\documentclass[preprint]{aastex631}
\usepackage{amsmath, bm}
\usepackage{graphicx}
\usepackage{xcolor}
\usepackage[normalem]{ulem} 
\newcommand\hl{\bgroup\markoverwith
  {\textcolor{yellow}{\rule[-.5ex]{2pt}{2.5ex}}}\ULon}
\shorttitle{AASTeX v6.3.1 Sample article}
\shortauthors{Ran et al.}
\graphicspath{{./}{figures/}}

\begin{document}

\title{Relationship between Successive Flares in the Same Active Region and Space-Weather HMI Active Region Patch (SHARP) Parameters}

\author{Hao Ran}
\affiliation{State Key Laboratory of Space Weather, National Space Science Center, Chinese Academy of Sciences, Beijing 100190, China; liuxying@swl.ac.cn}
\affiliation{University of Chinese Academy of Sciences, Beijing 100049, China}

\author{Ying D. Liu}
\affiliation{State Key Laboratory of Space Weather, National Space Science Center, Chinese Academy of Sciences, Beijing 100190, China; liuxying@swl.ac.cn}
\affiliation{University of Chinese Academy of Sciences, Beijing 100049, China}

\author{Yang Guo}
\affiliation{School of Astronomy and Space Science, Nanjing University, Nanjing 210023, China}

\author{Rui Wang}
\affiliation{State Key Laboratory of Space Weather, National Space Science Center, Chinese Academy of Sciences, Beijing 100190, China; liuxying@swl.ac.cn}


\begin{abstract}
A solar active region (AR) may produce multiple notable flares during its passage across the solar disk.
We investigate successive flares from flare-eruptive active regions, and explore their relationship with solar magnetic parameters.
We examine six ARs in this study, each with at least one major flare above X1.0.
The Space-Weather HMI Active Region Patch (SHARP) is employed in this study to parameterize the ARs.
We aim to identify the most flare-related SHARP parameters and lay foundation for future practical flare forecasts.
We first evaluate the correlation coefficients between the SHARP parameters and the successive flare production.
Then we adopt a Natural Gradient Boost (NGBoost) method to analyze the relationship between the SHARP parameters and the successive flare bursts.
Based on the correlation analysis and the importance distribution returned from NGBoost, we select 8 most flare-related SHARP parameters.
Finally, we discuss the physical meanings of the 8 selected parameters and their relationship with flare production.
\end{abstract}




\section{Introduction} \label{sec:intro}
As one of the most prominent eruptive phenomena, solar flares can produce significant space weather effects
including energetic particles.
In flare-eruptive active regions (ARs), flares may occur successively.
Since solar flares are often accompanied by coronal mass ejections (CMEs), successive flare bursts may appear with successive
CMEs, the interaction between which may result in extreme space weather \citep{Liustorm_nature, Liu_2019}.
Establishing a practical flare forecast system, especially for successive flare bursts in the same AR, is thus of importance for space weather.
Since the vector magnetic field can only be mapped directly in the photosphere, using photospheric magnetic field data to 
parameterize ARs has been widely adopted in studies about flare prediction\citep[e.g.,][]{yang2013SVM, bobra2015solar, wang2019parameters,wang2020solar, cicogna2021flare,yi2021visual,ribeiro2021machine}.
Due to the complex current distribution \citep{torok2014distribution} and the
magnetic field evolution, it is difficult to derive a single parameter 
that governs flare production in the same AR.
Therefore, identifying parameters that are the most relevant to flare productivity is of great importance.

The \emph{Helioseismic and Magnetic Imager} \citep[HMI; ][]{hmischou}, aboard the \emph{Solar Dynamics Observatory} (SDO),
continuously maps the full-disk photospheric vector magnetic field since 2010 May 1 every 12 minutes.
It also provides a data product named Space-Weather HMI Active Region Patch (SHARP), which offers a series of 
key parameters derived from the magnetic field in ARs \citep{bobra2014helioseismic,hoeksema2014helioseismic}.
\cite{2018Flare} applied $I_{8d}$, the sum of the GOES soft X-ray peak fluxes of flares 
in the same AR for 8 days, to investigate the correlation between SHARP parameters and flare productivity
in time-series, concluding that the SHARP parameters are to some extent
related to flare productivity. 
\cite{wang2019parameters} calculated the SHARP parameters in the polarity inversion line area of an AR, and 
employed a Support Vector Machine (SVM) algorithm to perform flare prediction with the modified SHARP parameters.
Many of the previous flare prediction studies focus on determining if an AR is flare-eruptive or flare-quiet,
instead of looking into the successive flare bursts along with the evolution of the AR \citep[e.g.,][]{yang2013SVM, bobra2015solar, wang2019parameters, cicogna2021flare,yi2021visual,ribeiro2021machine}.
Some studies considered the successive flare production, but focused on only limited parameters such as
the degree of current neutralization \citep[e.g.,][]{kontogiannis2017non}.
Here our approach is different.
We consider all the flares above C1.0 in the same AR, and study the relationship between the
successive flare bursts and SHARP parameters.

The polarity inversion line (PIL), where the polarity of the magnetic field is reversed, has been proved to have strong relationships with solar flares and CMEs \citep[e.g.,][]{2002Correlation, 2010TESTING, B2011Photospheric, 2012Non, 2012The, torok2014distribution, Liu2017Predicting, Wang_2018, Wang_2022}.
Parameters, such as the shear angle and the emerging magnetic flux, have been proved to have a 
proportional relationship with flare productivity, if narrowed to the PIL area \citep{1987Flare,2007Aschrijver}.
Previous works suggest that the PIL modified SHARP parameters perform better than original SHARP parameters 
with respect to flare prediction\citep[e.g.,][]{wang2019parameters,cicogna2021flare}.
In our study, we also employ similar methods to see the effect of PIL modification when studying the successive flares in the same ARs.

In this paper, we evaluate the relationship between SHARP parameters and successive flare production in two ways. 
The first method is the calculation of correlation coefficients.
The second method is based on a machine learning algorithm named Natural Gradient Boost (NGBoost), which is a 
gradient boost algorithm that can be easily used without time-consuming hyperparameter tuning \citep{2019NGBoost}.
This algorithm returns a probability distribution for the target and an importance distribution of the input parameters.
In this study, the input parameters are the SHARP parameters.
With the correlation result and the importance distribution returned from NGBoost, we could identify the most flare-related SHARP parameters, and make a 
physical analysis.

The paper is organized as follows.
In Section 2, we describe the data and the methodology that are employed in this paper. 
In Section 3, we show the correlation results, the importance distribution of the SHARP parameters, 
and the test of the selection. 
The underlying physics analysis of the selected parameters is also discussed in Section 3.
The conclusion is given in Section 4.

\section{Data and Methodology} \label{sec:data}
\subsection{Active Region Selection}
We have selected 6 ARs, which all have at least one major flare above level X1.0. 
Table \ref{tab:ARlist} lists the index, start and end time, number of flares, 
maximum flare, and the number of groups of SHARP data that we have employed for the ARs.
\begin{table}
    \centering
    \caption{List of active regions in our study.}
    \begin{tabular}{c c c c c c} \hline
        AR & Start time & End time & Number of flares & Maximum flare & Groups of data \\ \hline
        11158 & 2011-02-13 11:58:12 & 2011-02-18 11:46:12 & 41 & X2.2  &  600  \\
        11166 & 2011-03-08 23:58:14 & 2011-03-11 23:46:15 & 12 & X1.5  &  357  \\ 
        11283 & 2011-09-05 23:58:22 & 2011-09-10 23:46:21 & 13 & X2.1  &  599  \\
        11429 & 2012-03-04 15:58:14 & 2012-03-12 13:58:15 & 45 & X5.4  &  900 \\
        12017 & 2014-03-26 23:58:16 & 2014-03-31 23:46:17 & 15 & X1.0  &  588  \\
        12673 & 2017-09-01 23:58:42 & 2017-09-07 18:46:41 & 42 & X9.3  &  620  \\
        \hline
    \end{tabular}
    \label{tab:ARlist}
\end{table}

Since the ARs are at different latitudes, and the longitudes of them are also different when they are formed, the time durations for them to stay on the 
visible side of the Sun are different, resulting in differences in the effective observation times.

The ARs have been studied widely from many aspects, such as the energy release process \citep[e.g.,][]{2013lucasAR11158, Feng_2013_ar11283, 2014wangrui11429, Wang_2018, Wang_2022}, 
the structure of the nonlinear force-free field \citep[e.g.,][]{Feng_2013_ar11283, Wang_2013, 2014vemareddy11166},
and the CMEs associated with the major flares \citep[e.g.,][]{2013liuapj11423, 2014LiuApJLar11429}.

\subsection{Flare ``Envelope" Curve}
\begin{figure}
    \centering
    \includegraphics[width=0.95\textwidth, height=0.45\textwidth]{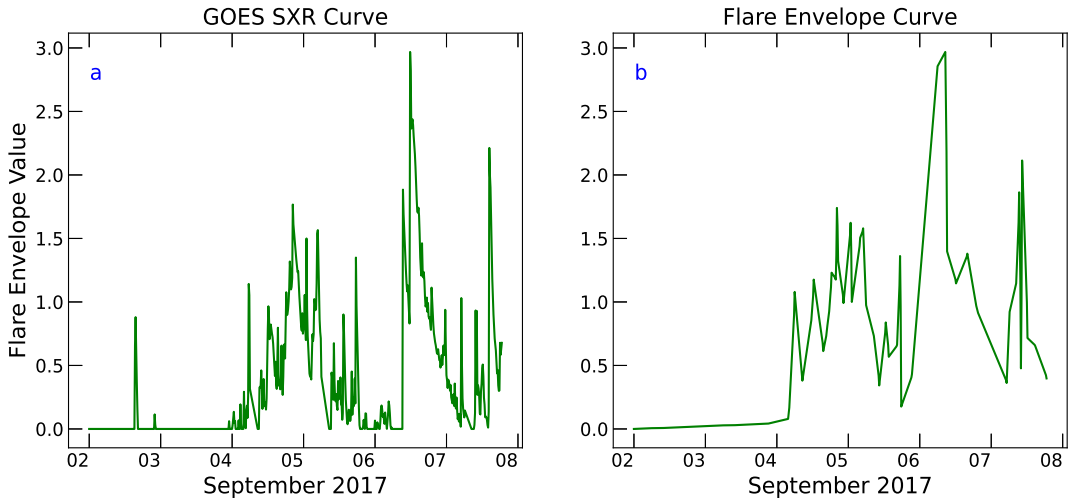}
    \caption{The original GOES SXR curve and the flare ``envelope'' of AR 12673. Panel ``a'' refers to the original GOES SXR curve, and panel ``b''
    represents the corresponding flare ``envelope''. The temporal resolution of both panels is 720s, which is the same as 
    that of the SHARP data.}
    \label{fig:envelope}
\end{figure}

The intensity of flares is defined by GOES Soft X-ray (SXR) flux, and we only select the flares bigger than C1.0
that has a peak flux of $1 \times 10^{-6} \  W/m^2$. We ignore the temporal evolution of an individual flare, and define 
its time tag as the flare peak time. The curve ``enveloping" successive flares is calculated from:
\begin{equation}
    f_i = log_{10}(\Phi_i) + 6,
\end{equation}
where $f_i$ is the defined flare index in the flare ``envelope'', and $\Phi_i$ is the corresponding peak flux.
The flare data and the GOES SXR data are obtained from the SolarflareDatabase \footnote{https://solarflare.njit.edu/datasources.html}.
An example of the flare ``envelope'' of AR 12673 is shown in Figure \ref{fig:envelope}b. 

The use of the flare envelope allows us to derive the flare data in a specific
AR instead of the whole solar disk, and set a threshold for the flares conveniently.
Flares from other ARs can be readily eliminated.
It also allows us to easily adjust the temporal resolution of the flare curve to 
match the temporal resolution of SHARP. 
The main disadvantage of the envelope curve is that it is obtained from discrete flare 
bursts. Temporal details of individual flares are ignored when the flare ``envelope'' 
curve is employed.

Note that the flare envelope curve is less zigzag than the curves derived 
from the original GOES SXR flux curve.
For example, we have tried another curve that is directly derived from the GOES SXR flux curve (see Figure \ref{fig:envelope}a). 
The GOES SXR data points are kept if they are within the timespan of the flares from the selected AR.
The removed data points are filled by a linear interpolation.
Compared to the flare envelope curve, the new curve is not significantly different in shape but much more zigzag.
A more zigzag curve of flares will increase the difficulty for correlation analysis and further NGBoost analysis.

\subsection{Space-Weather HMI Active Region Patch (SHARP)}
SHARP\footnote{http://jsoc.standford.edu/ajax/lookdata.html} provides a series of key parameters derived from photospheric magnetic fields.
We have selected 16 typical SHARP parameters \citep{bobra2014helioseismic}.
Table \ref{tab:SHARP} lists the keywords, the descriptions and the calculation formulas for all the SHARP parameters we have used in this study.
The parameters are generally adopted from \cite{2003Photospheric}.
\begin{table}
    \centering
    \caption{Brief descriptions and corresponding formulas of the SHARP parameters \citep{bobra2014helioseismic}.}
    \begin{tabular}{ c c c }\hline
        Keyword & Description & Calculation formula \\ \hline
        TOTUSJH & Total unsigned current helicity & $ \bm{H}_{c_{total}} \propto \sum | \bm{B_z} \cdot \bm{J_z} | $ \\
        TOTPOT & Total photospheric magnetic free energy density & $ \rho_{tot} \propto \sum (\bm{B}^{obs} - \bm{B}^{pot})^2 dA $ \\
        TOTUSJZ & Total unsigned vertical current & $ \bm{J}_{z_{total}} = \sum | \bm{J}_z | dA $ \\
        ABSNJZH & Absolute value of net current helicity & $ \bm{H}_{c_{abs}} \propto | \sum \bm{B}_z \cdot \bm{J}_z | $ \\
        SAVNCPP & Sum of modulus of the net current per polarity & $ \bm{J}_{z_{sum}} \propto | \sum_{B_z^{+}} \bm{J}_z dA| + | \sum_{B_z^{-}}  \bm{J}_z dA | $ \\
        USFLUX & Total unsigned flux & $\Phi = \sum | \bm{B}_z dA |  $ \\
        MEANPOT & Mean photospheric magnetic free energy & $ \overline{\rho} \propto \frac{1}{N} \sum ( \bm{B}^{obs} - \bm{B}^{pot})^2  $ \\
        R\_VALUE & Sum of flux near PIL & $ \Phi = \sum |\bm{B}_z| dA $ within R mask  \\
        MEANSHR & Mean shear angle & $ \overline{\Gamma} = \frac{1}{N} \sum cos^{-1} ( \frac{ \bm{B}^{obs} \cdot \bm{B}^{pot}}{|\bm{B}^{obs}| |\bm{B}^{pot}|}) $ \\
        MEANGAM & Mean angle of field from radial & $ \overline{\gamma} = \frac{1}{N} \sum tan ^ {-1} \frac{B_h}{B_z} $ \\
        MEANGBT & Mean gradient of total field & $ \overline{|\nabla \bm{B}_{tot}|} = \frac{1}{N} \sum [ (\frac{ \partial B_{tot} }{ \partial x})^2 + (\frac{ \partial B_{tot}}{\partial y})^2]^{(1/2)} $ \\
        MEANGBZ & Mean gradient of vertical field & $ \overline{|\nabla \bm{B}_z|} = \frac{1}{N} \sum [ (\frac{\partial B_z}{\partial x})^2 + (\frac{\partial B_z}{\partial y})^2]^{(1/2)} $ \\
        MEANGBH & Mean gradient of horizontal field & $ \overline{|\nabla \bm{B}_h|} = \frac{1}{N} \sum [ (\frac{\partial B_h}{\partial x})^2 + (\frac{\partial B_h}{\partial y})^2]^{(1/2)} $ \\
        MEANJZH & Mean current helicity ($B_z$ contribution) & $ \overline{\bm{H}_c} \propto \frac{1}{N} \sum \bm{B}_z \cdot \bm{J}_z $ \\
        MEANJZD & Mean vertical current density & $ \overline{\bm{J}_z} \propto \frac{1}{N} \sum (\frac{\partial B_y}{\partial x} - \frac{\partial B_x}{\partial y})  $ \\
        MEANALP & Mean characteristic twist parameter: $\alpha$ & $ \overline{\alpha} \propto \frac{\sum \bm{J}_z \cdot \bm{B}_z }{ \sum \bm{B}_z^2} $ \\
        \hline
    \end{tabular}
    \label{tab:SHARP}
\end{table}

\subsection{Correlation and NGBoost Analysis}
First, we look at the SHARP parameters for the whole AR, and calculate the correlation 
coefficients between the SHARP curves and the flare envelope. 
We also look at the SHARP parameters in the PIL area to study how PIL modification 
affects the correlation results.
The method to identify the PIL initially comes from \cite{2007Aschrijver}.
To be specific, we set a threshold of $2 \sigma$ (200G) \citep{hoeksema2014helioseismic}, 
any positive field pixel with $B_r > 200$G is set to 1 and the rest are 0 to make the 
positive bitmap,
and any negative field pixel with $ B_r < -200$G is set to -1 and the rest are 0 to make the 
negative bitmap. 
Then we adopt a cluster algorithm called Density-Based Spatial Clustering of Applications with Noise \citep[DBSCAN; ][]{ester1996density} to 
remove the noise data points and leave only the main parts of the positive and negative 
bitmaps.
We finally multiply the two bitmaps after convolving them with a Gaussian 
kernel to generate the PIL mask. 

We also adopt a Natural Gradient Boost (NGBoost) method\footnote{https://stanfordmlgroup.github.io/projects/ngboost} \citep{2019NGBoost} to analyze the relationship between flare production and SHARP parameters 
and obtain the importance distribution. The NGBoost method is a gradient boosting algorithm for probabilistic regression problems, 
which returns a possibility distribution of the target.
The possibility distribution is defined by two parameters, ``location" and ``scale". 
The ``location" parameter represents the location of the maximum point of the distribution. It shifts the graph, relative to the 
normal probability distribution function (PDF), left or right on the horizontal axis depending on if its value is negative or positive.
The ``scale" parameter defines the width of the distribution. It expands or shrinks the graph, relative to the normal PDF,
depending on whether its value is larger or smaller than 1.

We apply the NGBoost method to SHARP parameters as follows.
We discard the first 2 hours of flare envelope data, and match the remaining flare envelope data to the original SHARP parameters, 
in order to establish a 2-hour time difference between the flare envelope and the SHARP curve.
This is to explore whether the SHARP parameters before the flares can reflect flare productivity.
Then, we take the first 10 groups of data, which includes the first 10 groups of SHARP parameters and the first 10 flare envelope points, to be the initial training set.
We start the calculation from the 11th groups of the SHARP parameters. 
With the training of the first 10 groups of data finished, we input the feature vector of the 11th group of SHARP parameters to the trained model, and obtain the 
corresponding ``flare envelope'' point calculated from NGBoost.
An importance distribution of the SHARP parameters in this training step is returned from NGBoost as well.
In the next step, we take the first 11 groups of data as the training set, input the 12th SHARP parameters to the trained model, 
and obtain the next calculated ``flare envelope'' point along with a new importance distribution.
This operation is conducted to the last group of data.
Finally, we attain a new ``flare envelope'' calculated from NGBoost and the importance distributions for all the steps.
The final importance distribution of each SHARP parameter is the average value of all the importance distributions that we have obtained.
The explanation of the NGBoost method can be found in the Appendix.
It is worth mentioning that the 2-hour difference between the flare envelope and the SHARP parameters is only used in the ``NGBoost Analysis'' section (Section 3.2).
In the other two sections of Section 3, the SHARP parameters and the flare envelope are simultaneous.

\section{Results}
\subsection{Correlation Analysis}
\begin{figure}
    \centering
    \includegraphics[width=1.0\textwidth,height=0.75\textwidth]{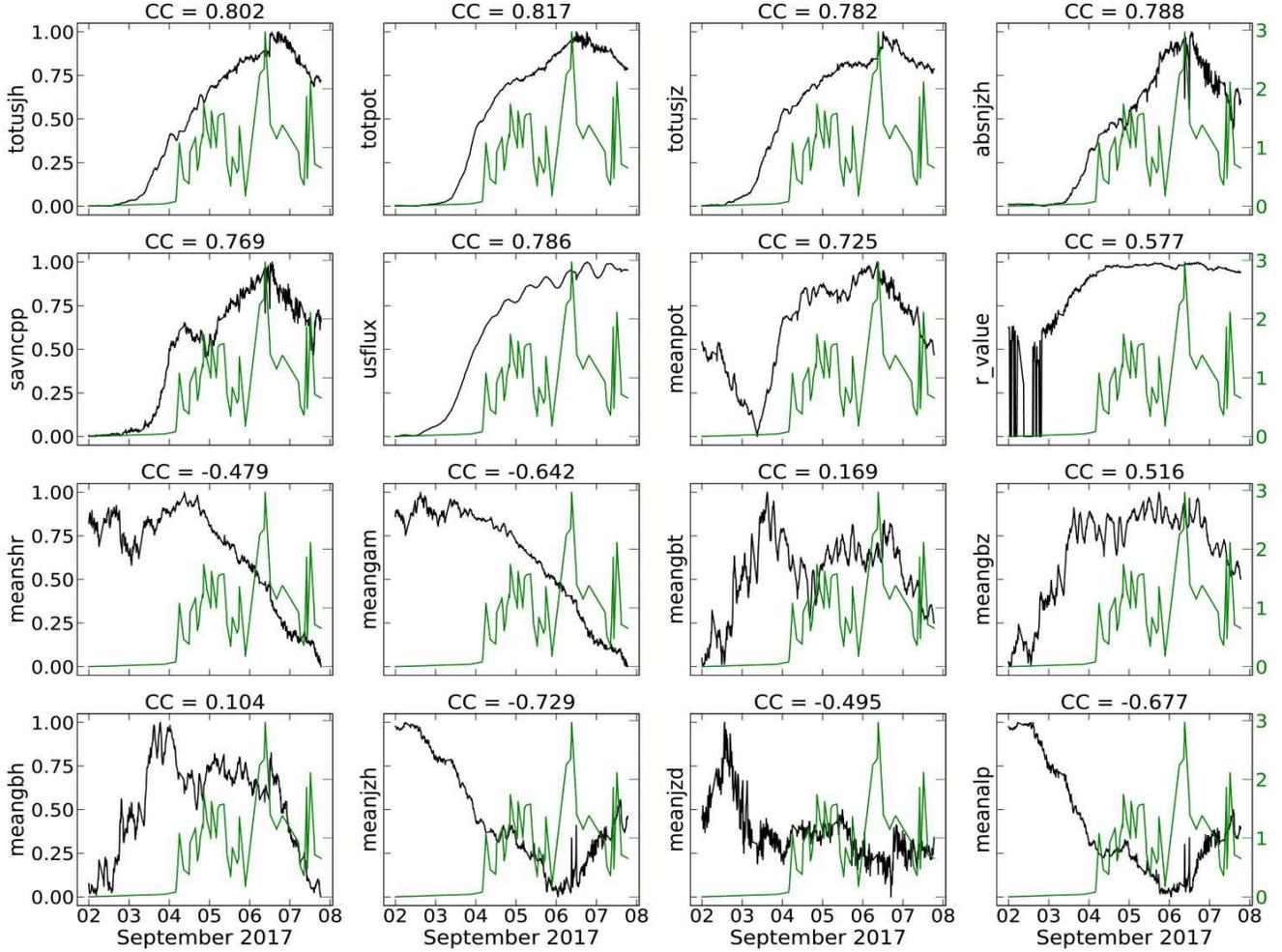}
    \caption{The 16 SHARP parameters and the flare envelope for AR 12673. The black curve shows different 
    SHARP parameters and the green curve is the flare envelope. The left y-axises represent the values of the SHARP 
    parameters, and the right y-axises represent the flare index. The correlation coefficients (CCs) are shown at the top 
    of each panel. The SHARP parameters are normalized.}
    \label{fig:AR12673}
\end{figure}

As mentioned above, we first look at the SHARP parameters in the whole AR and calculate their 
correlation coefficients with the flare envelope to estimate how 
each SHARP parameter is related to the flare envelope. 
We take the AR 12673 as an example in Figure \ref{fig:AR12673}.
As can be seen in Figure \ref{fig:AR12673}, in AR 12673, the range of the correlation coefficients (CCs) is wide for different SHARP parameters.
For example, the CC for ``TOTPOT'' is over 0.8, while the absolute value of the CC for ``MEANGBH'' is only 0.104, 
which suggests that ``MEANGBH'' is much less relevant to the flare envelope than ``TOTPOT'' in this AR.

\begin{figure}
    \centering
    \includegraphics[width=1.0\textwidth,height=0.4\textwidth]{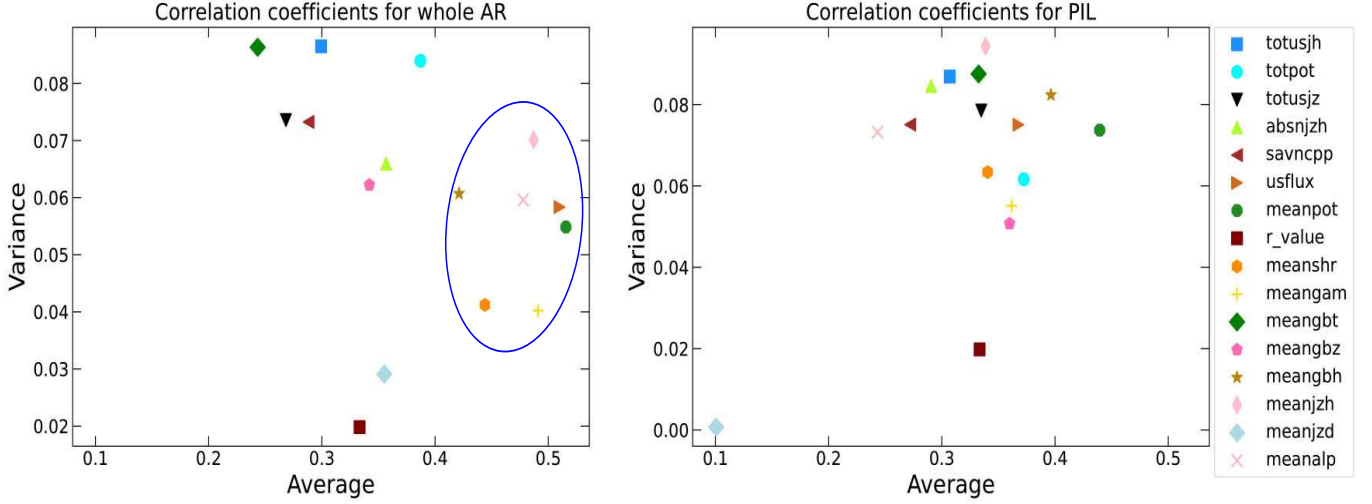}
    \caption{Scatter plot of the average and variance of the correlation 
    coefficients between the SHARP parameters and the flare envelope. The left panel refers to 
    the whole AR case, and the right panel refers to the PIL-modified case.
    The x-axis represents the average value, and y-axis represents the variance value. 
    A parameter has a better performance if it is closer to the lower right corner 
    of the figure. Inside the blue ellipse are the parameters that we think have the best 
    performance in correlation.}
    \label{fig:scattercorrelation}
\end{figure}

We calculate the CCs for all the SHARP parameters, in the whole AR and in the PIL area, for all the ARs that we have selected.
Then we plot the average and variance of the CCs for SHARP parameters in Figure \ref{fig:scattercorrelation}.
Figure \ref{fig:scattercorrelation} consists of two panels, referring to the SHARP parameters in the whole AR and the SHARP parameters in the PIL area.
We mark the parameters that we think have the best performance in correlation with the blue ellipse in the left panel, 
which includes ``MEANJZH'', ``MEANGBH'', ``MEANALP'', ``USFLUX'', ``MEANPOT'', ``MEANSHR'' and ``MEANGAM''.

\begin{figure}
    \centering
    \includegraphics[width=1.0\textwidth, height=0.7\textwidth]{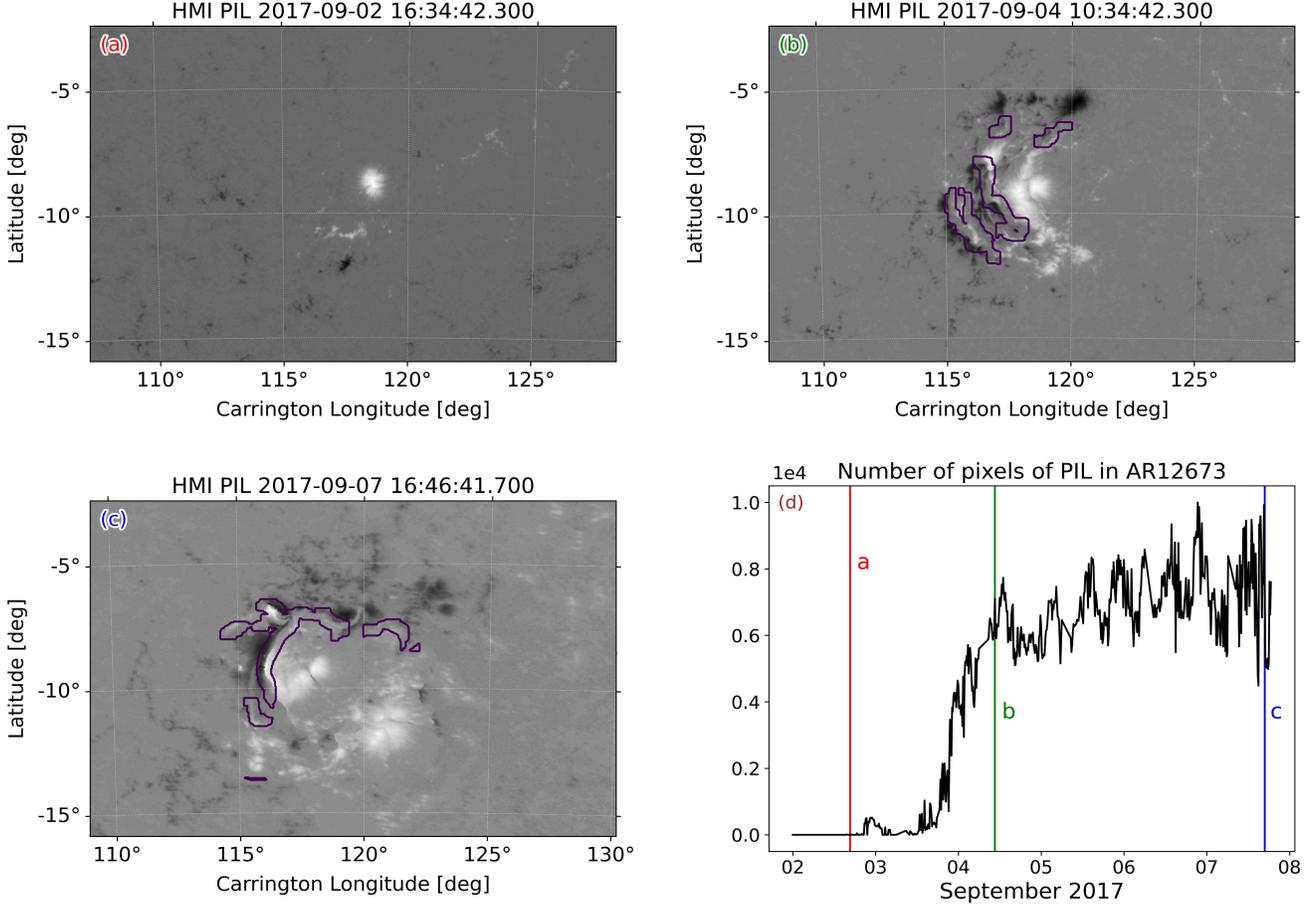}
    \caption{Illustration of the change in the coverage and the number of PILs in AR 12673. 
    Panel (a), panel (b) and panel (c) are the composite maps of the Br map and their PILs at different times.
    The PIL area is marked by the purple lines.
    Panel (d) shows the change in the number of pixels of PILs in AR 12673.
    The three vertical lines in panel (d) correspond to the times of panels (a-c).}
    \label{fig:pil_change}
\end{figure}

Our results in Section 3.3 present that, in agreement with previous studies \citep[e.g.,][]{2007Aschrijver, wang2019parameters}, the SHARP parameters that are believed to be more related to successive flare occurrence do show more significant
changes in the PIL area when major flares happen.
However, the two panels in Figure \ref{fig:scattercorrelation} suggest that the average values of some 
SHARP parameters in the PIL area, such as ``MEANPOT'' and ``MEANJZH'', are lower than those in the whole AR.
This is due to sharper fluctuations of the SHARP curves in the PIL area.
The value of the correlation coefficient between the two curves is not only affected by their overall shapes, but also their smoothness.
Therefore, the SHARP curves in the PIL area may fit the flare envelope better in the overall shape, but the correlation coefficients may be lower due to their stronger oscillation.
The more intense fluctuations of the SHARP curves in the PIL area can be attributed to two reasons:
(i) During the evolution of a big AR, the coverage of the PIL area varies, which results
in fluctuations in SHARP parameters calculated in the PIL area. 
(ii) Some ARs have more than one PIL at one time, while some other ARs may have no obvious PIL. 
Even in the same AR, such as AR 12673, the quantity of PILs may
change with evolution.
Figure \ref{fig:pil_change} presents the evolution of the PILs in AR 12673.
AR 12673 starts with no obvious PIL (see Figure \ref{fig:pil_change}a).
However, as it becomes more active, it has more than one PIL and the coverage of the PILs increases significantly (see Figure \ref{fig:pil_change} (b-c)).
Panel ``d'' in Figure \ref{fig:pil_change} also shows that the coverage of the PILs fluctuates strongly 
when the AR gets complicated, which may cause severe fluctuations in the SHARP curve.
Similar features about the change in the coverage and the number of PILs can be found in all the ARs that we select.

\subsection{NGBoost Analysis}
\begin{figure}
    \centering
    \includegraphics[width=1.0\textwidth, height=0.55\textwidth]{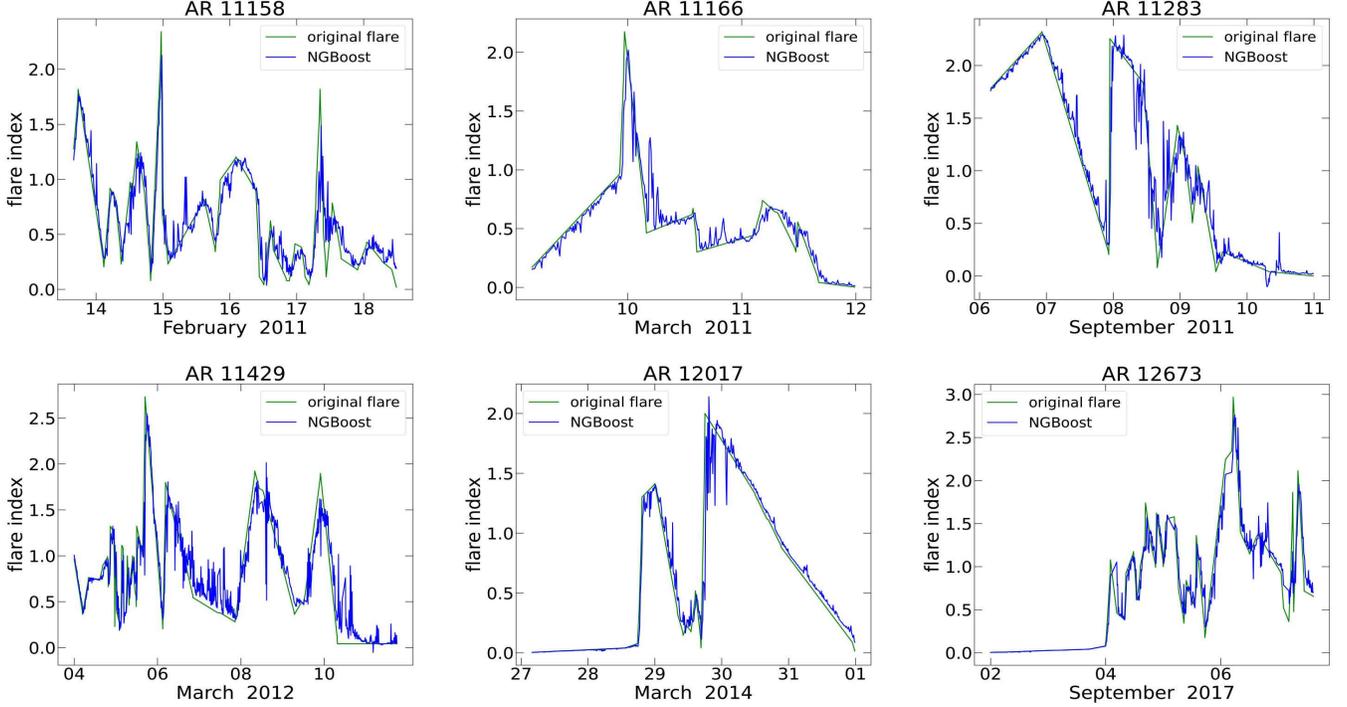}
    \caption{The original flare envelope and the corresponding ``flare envelope'' calculated from NGBoost for the
     6 selected ARs. The green curves are the
     original flare envelopes, and the blue curves are from NGBoost.}
    \label{fig:allNGBoost}
\end{figure}

Since the correlation coefficients between the flare envelope and the SHARP in the whole AR are usually higher, 
we only apply the NGBoost method to the SHARP parameters in the whole AR.
The result is shown in Figure \ref{fig:allNGBoost}. The blue curve is the ``flare envelope'' we calculated from NGBoost, and the green curve 
is the original flare envelope. The NGBoost curves show good fits to the flare envelopes in all the 
ARs we select.

\begin{figure}
    \centering
    \includegraphics[width=1.0\textwidth,height=0.35\textwidth]{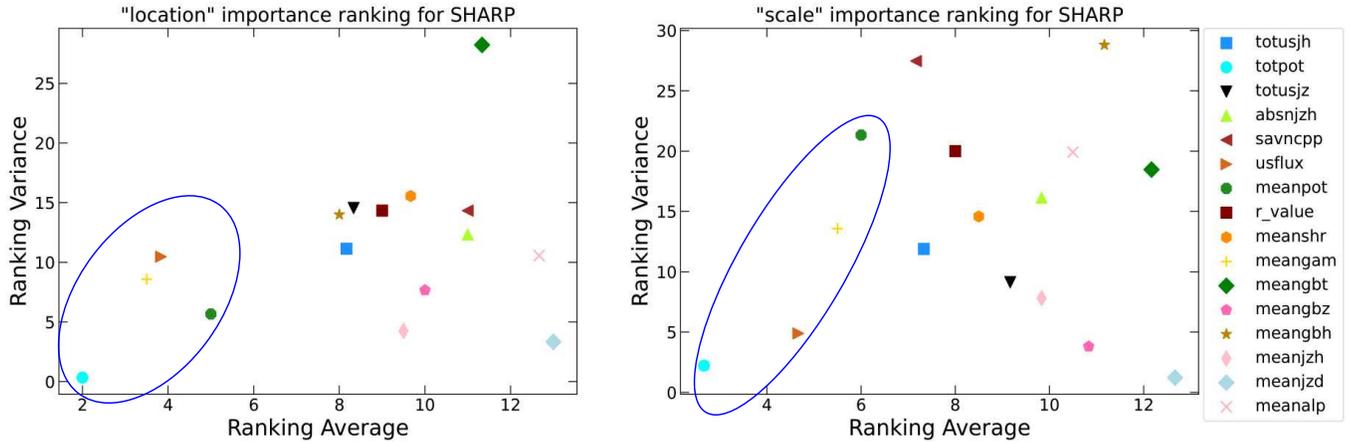}
    \caption{Scatter diagrams for the ranking of different SHARP parameters calculated from NGBoost. 
    The left panel shows the ranking for the parameter ``location'' in the possibility distribution, 
    and the right one shows the ranking for the parameter ``scale''. 
    Inside the blue ellipse are the parameters of the greatest importance.}
    \label{fig:scatterNGBoost}
\end{figure}

With the importance distributions of the SHARP parameters returned from NGBoost, 
we calculate the ranking of different SHARP parameters as shown in Figure \ref{fig:scatterNGBoost}.
It contains the ranking average and corresponding variance of all the SHARP parameters in 
the 6 ARs. The x-axis stands for the ranking average and the y-axis stands for the
ranking variance, both of which are better when smaller. 
Thus, if a parameter is closer to the bottom left corner, it is of more importance. 
As Figure \ref{fig:scatterNGBoost} suggests, we can derive ``TOTPOT'', ``MEANPOT'', ``USFLUX'' and  ``MEANGAM'' 
out of the 16 parameters as the good performance group. 
Combining the correlation analysis in Figure \ref{fig:scattercorrelation}, we put another 4 SHARP 
parameters into the good performance group.
The 4 SHARP parameters are ``MEANJZH'', ``MEANGBH'', ``MEANALP'', ``MEANSHR''.
All the selected parameters have been marked by the blue ellipses.

\begin{figure}
    \centering
    \includegraphics[width=1.0\textwidth, height=0.55\textwidth]{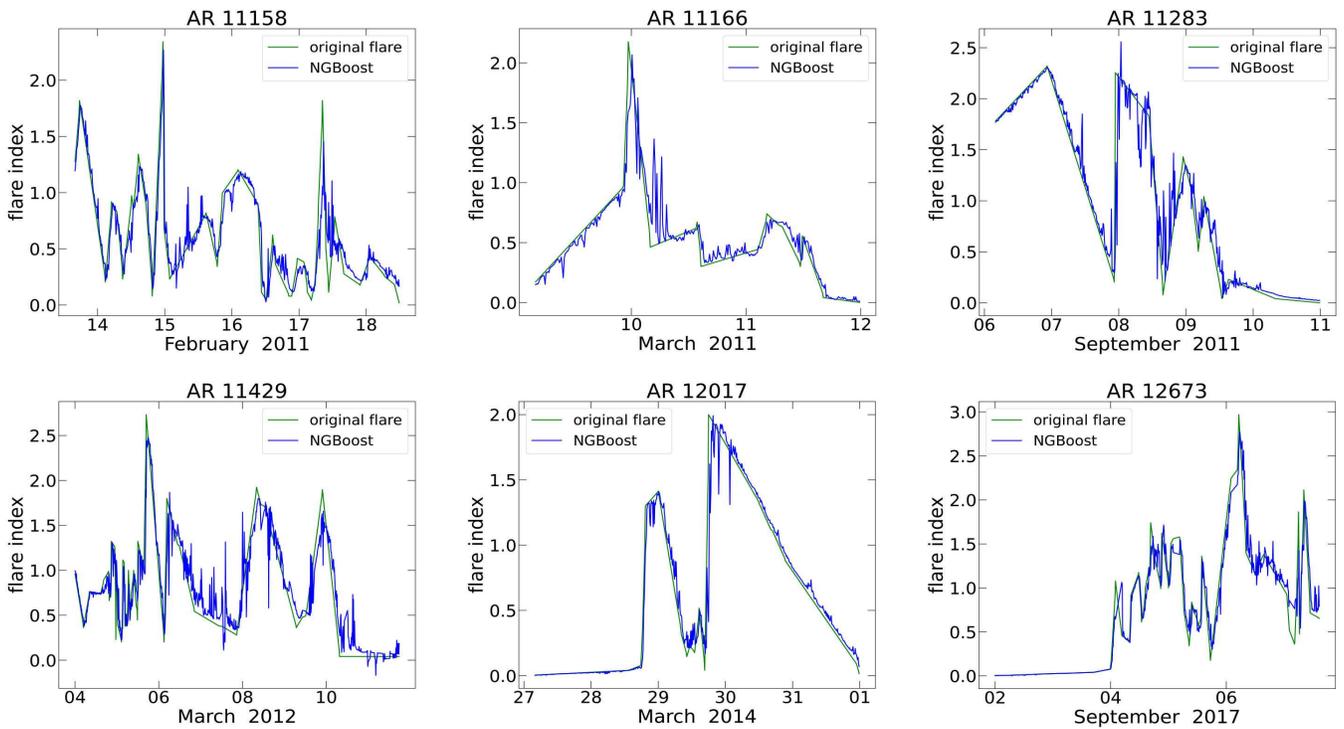}
    \caption{Comparison between the original flare envelope and the ``flare envelope'' curve calculated from NGBoost with 
    the selected 8 SHARP parameters only.}
    \label{fig:6NGBoost}
\end{figure}

Before we start to analyze the 8 parameters separately, we have verified the reliability of 
the 8 parameters by applying the NGBoost method once more to the 8 parameters 
instead of all the SHARP parameters.
The result is shown in Figure \ref{fig:6NGBoost}.
We have also conducted the same operation with fewer SHARP parameters, 
and the fitting of the two curves is not as good as the eight-element result.
This suggests that the selection is successful and convincing.

\subsection{Underlying physics of the selected SHARP parameters}

\begin{figure}
    \centering
    \includegraphics[width=1.0\textwidth, height=0.8\textwidth]{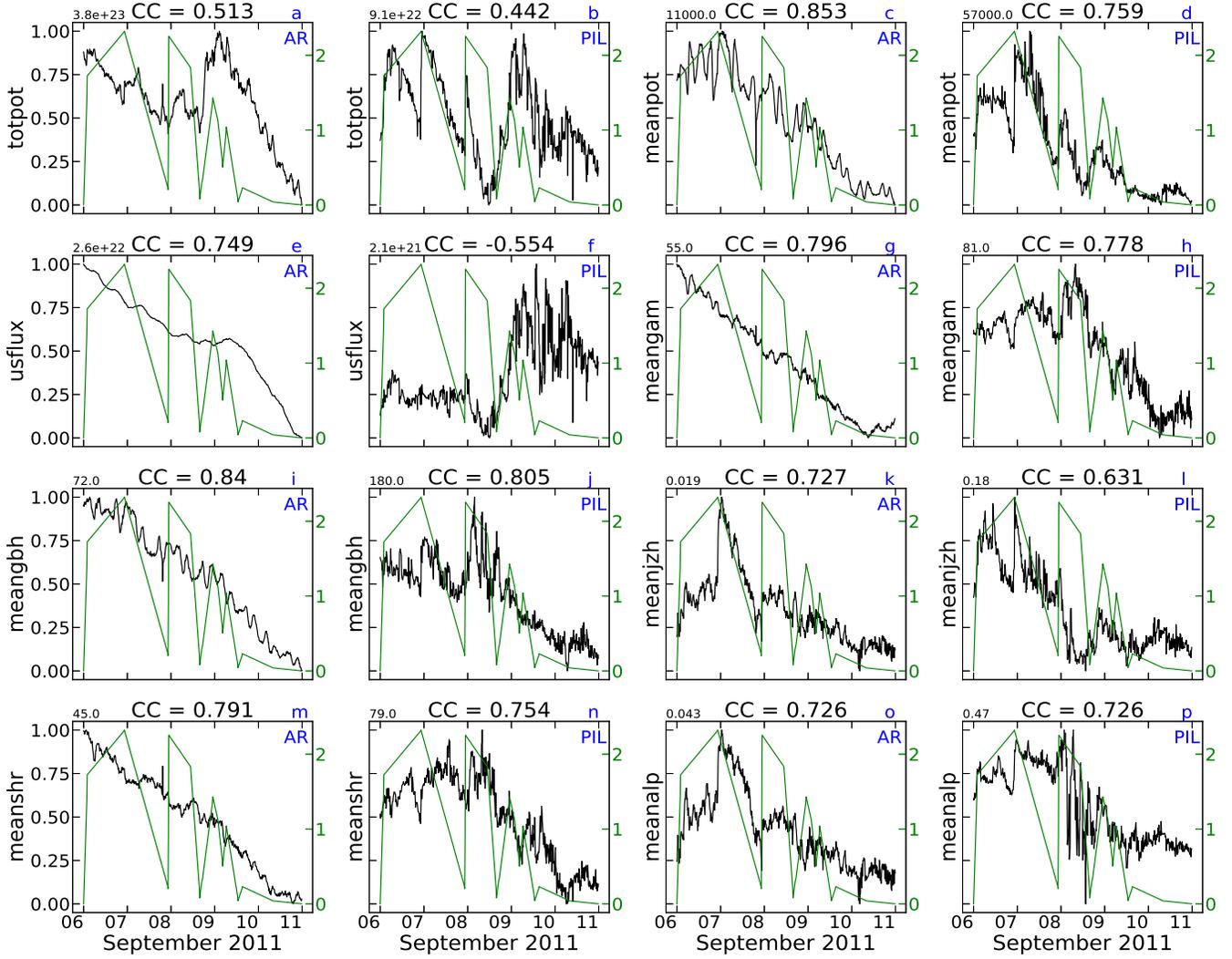}
    \caption{The 8 parameters in the whole AR and in the PIL area for AR 11283. 
    Two adjacent panels, such as panel ``a'' and ``b'', belong to a group of the same SHARP.
    In the same group, the left panel is the SHARP in the whole AR, and the right one is the SHARP in the PIL area.
    Counting from left to right and up to down, the 8 groups are: ``TOTPOT'', ``MEANPOT'', ``USFLUX'', ``MEANGAM'', ``MEANGBH'', ``MEANJZH'', ``MEANSHR'' and ``MEANALP''.
    Each panel is labeled with a letter from ``a'' to ``p''. The SHARP parameters are normalized.
    The corresponding correlation coefficient (CC) is at the top of each panel.
    The maximum value of each SHARP parameter is also at the top. 
    Similar format to Figure \ref{fig:AR12673}.
    }
    \label{fig:AR11283}
\end{figure}

We analyze the 8 selected SHARP parameters in this section.
Figure \ref{fig:AR11283} shows an example for the evolution of the selected SHARP parameters in the whole AR and the PIL area for AR 11283.
Two adjacent panels, such as the panel ``a'' and panel ``b'', represent the same parameter. 
The left panel shows the SHARP in the whole AR, and the right one shows the SHARP in the PIL area.
Here we include the PIL-modified SHARP parameters to study how the SHARP parameters evolve in the PIL area during flares.

The first two parameters are ``TOTPOT'' and ``MEANPOT'', which have similar
physical meanings but are derived in different ways. ``TOTPOT'' is the total value of the magnetic 
free energy, and ``MEANPOT'' is the average value. Free energy is highly related to the 
nonlinear-force-free field. According to previous studies, a great amount of accumulated 
magnetic free energy is released within a short time to provide the energy for flares and CMEs 
\citep[e.g.,][]{antiochos1999model,amari1999twisted,chen2000emerging,2014wangrui11429}.
\cite{guo20083d} analyzed the flare in December 2006 by extrapolating its 3D magnetic field 
configuration. They calculated the free energy and the shear angle of the AR, 
studied the energy release process throughout the evolution of the magnetic field, and 
explained the energy release process during a flare. \cite{schrijver2008nonlinear} also 
explained the free energy release process by extrapolating the nonlinear-force-free field, and 
found that a free energy drop of $\sim 10^{32}$ erg happened as the X flare occurred. 
In our study, the two SHARP parameters about the free energy are also considered as two of the 
most related parameters for flare occurrence. 

In Figure \ref{fig:AR11283}, panel ``a'' and panel ``b'' represent ``TOTPOT'', and panel ``c'' and panel ``d''
represent ``MEANPOT''. We can see in the panels that, comparing to the two parameters in the whole AR, ``TOTPOT''
and ``MEANPOT'' show a more obvious increase in the PIL area when big flares happen. 
The absolute value of ``MEANPOT'' is much higher than that in the whole AR.
These features are found in all the ARs we select, which suggests that the storing and releasing process of free energy is more obvious in the PIL area.
This conclusion is in agreement with previous studies  \citep[e.g.,][]{cicogna2021flare,wang2019parameters}.

The next parameter that we look into is ``USFLUX", which represents the total unsigned flux.
``USFLUX'' is shown in panels (e-f) in Figure \ref{fig:AR11283}.
The correlation between the total unsigned flux and flare productivity has been revealed by previous studies \citep[e.g.,][]{2007Aschrijver, vasantharaju2018statistical, bobra2015solar, kazachenko2017database}. 
\cite{vasantharaju2018statistical} made a statistical study of some magnetic 
imprints and their correspondence with flare productivity.
Following the results of previous studies \citep[e.g.,][]{2007Aschrijver,kazachenko2017database}, they took three kinds of total unsigned magnetic flux into consideration, 
the flux of the whole AR, the flux along the PIL, and the flux in the flare ribbon area. 
Comparing to the total unsigned flux in the whole AR, a slight and a strong increase in 
correlation with flares are found along with the PIL and in the flare ribbon area, respectively.

The two parameters of ``MEANGAM" and ``MEANGBH" are both related to the horizontal magnetic field. 
In Figure \ref{fig:AR11283}, panel ``g'' and panel ``h'' represent ``MEANGAM'', and panel ``i'' and panel ``j'' represent ``MEANGBH''.
The main finding about the horizontal magnetic field is that the horizontal field tends to increase 
significantly in the PIL area after flare eruptions, indicating that the 
vector magnetic field tends to be more horizontal after flares 
\citep{petrie2012abrupt,petrie2012spatio,sun2012evolution, 2014wangrui11429}.
This is consistent with our conclusion drawn from Figure \ref{fig:AR11283}.
Compared to panel ``g'' and panel ``i'', the SHARP curves in panel ``h'' and panel ``j'' show more 
obvious increases in the PIL area when big flares happen. This feature of these two parameters is also found in all the ARs we select.
Previous studies suggest that a rapid and irreversible increase appears in the
horizontal magnetic field after an eruption, and it becomes more aligned with the PIL as well.
\cite{song2016relationship} performed a detailed study by considering the continuum intensity. 
They compared the continuum intensity and the horizontal magnetic field in 5 ARs, and 
revealed that the differences of the continuum intensity and the horizontal magnetic field before 
and after the major flares (above X1.0) are strongly correlated both spatially and temporally 
in the outer penumbra of the sunspot. 
A similar conclusion has been drawn by \cite{lu2019statistical} through a
statistical analysis of 20 X-class flares. 
All these works indicate that the horizontal magnetic field is of great importance for successive flares in the same AR.

The parameter "MEANJZH", shown in panels (k-l) in Figure \ref{fig:AR11283}, stands for the mean current helicity with $B_z$ contribution. 
Current helicity is crucial to magnetic energy build-up. 
An explanation for the relationship between flares and current helicity is the 
``$\alpha-$effect'', indicating that the energy of small-scale fluctuations of magnetic and 
velocity fields could be transferred into the energy of large-scale currents \citep{seehafer1990electric}. 
The necessary condition for such effect may be the presence of the predominant sign of the 
electric helicity over the whole AR. 
\cite{abramenko1996analysis} conducted a statistical study involving 40 ARs and 
suggested that the condition of the ``$\alpha-$effect'' worked in over $90\%$ of all the cases.
\cite{2008Evolutionhelicity} performed a simulation to study the current helicity. 
They discussed the sources of the current helicity and suggested that field lines are skewed as 
they cross the PIL. 
Similar to current helicity, magnetic helicity is also believed to be closely related 
to the mechanisms that drive solar activities. 
``MEANALP'', shown in panels ``o'' to ``p'' in Figure \ref{fig:AR11283}, represents the mean characteristic twist parameter.
The method used to calculate this SHARP parameter was introduced by \cite{Hagino_2004helicity} and adopted by \cite{bobra2014helioseismic} in the SHARP package.
\cite{Wang_2022} studied the 4 homologous events (all are eruptive flares) in AR 11283.
They analyzed the buildup process of the magnetic flux ropes (MFRs), especially the evolution of the magnetic helicity and the energy during the 4 eruptions.
Their results show that the MFRs, which are built up along the collisional polarity inversion lines (cPILs), store a large amount of energy and magnetic helicity.
The buildup process is involved with magnetic cancellation along the cPIL via collisional shearing, which may contribute to solar eruptions \citep{Chintzoglou_2019}.
\cite{Li_2022} proposed several new parameters and investigated their ability in distinguishing eruptive and confined solar flares.
They found that the ratio between the ``MEANALP'' in the flaring PILs and the total unsigned flux in the whole AR ($\alpha_{FPIL}/\Phi_{AR}$) is the most capable parameter to distinguish ARs that can produce eruptive flares.
In our work, as shown in Figure \ref{fig:AR11283}, the values of the parameters ``MEANJZH'' and ``MEANALP'' in the PIL area are about ten times the values in the whole AR, indicating that both current and magnetic helicities accumulate in the PIL area.
This is in agreement with the mentioned studies.

The last parameter we select is ``MEANSHR'', which stands for the mean shear angle.
The shear angle is defined as the angle between the potential field and the observed field in the photosphere.
It is commonly accepted that a more strongly sheared magnetic filed stores more free energy.
\cite{Kusano_2012}
conducted a simulation work to interpret the types of magnetic field structures that can drive solar eruptions.
Their simulation results show that the shear angles of large-scale eruptions accurately indicate the kinetic energy produced by eruptions.
More sheared magnetic fields favor the onset of large-scale eruptions.
A lot of studies have been published to reveal the relationship between shear angles and flare production, suggesting that the shear angle is important in studying flare bursts \citep[e.g.,][]{1987Flare, Zhang_2001, Leka_4, guo20083d, Wang_2022, Li_2022}.

\section{Summary and Conclusions}
This paper has identified the SHARP parameters that are most related to successive flare 
bursts in the same AR.
For this purpose, we adopt the envelope of the successive flare bursts 
to represent discrete individual events, and calculate the correlation coefficients between the 
envelope and the SHARP parameters. 
The NGBoost method \citep{2019NGBoost} is also employed to obtain the importance distribution of the SHARP parameters. 
With the importance distribution from NGBoost and the correlation results, we have selected 8 parameters.
A time difference of 2 hours between the SHARP parameters and the flare envelope is established in the NGBoost section, which is 
for exploring if the SHARP parameters before flares could reflect flare productivity.
In fact, we have conducted the NGBoost analysis 10 times with time differences varying from 12 minutes to 2 hours,
and the selections of SHARP parameters are almost the same.
By comparing our results with previous works, we analyze the physical meanings of the 8 parameters and their relationships with flare productivity.
The major conclusions are summarized below.
\begin{enumerate}
    \item The most relevant SHARP parameters to the successive flare productivity in the 
    same AR are ``TOTPOT'', ``MEANPOT'', ``USFLUX'', ``MEANGAM'', ``MEANGBH'', ``MEANJZH'', ``MEANSHR'', and ``MENAALP''. 
    These parameters are related to the free energy, unsigned magnetic flux, horizontal magnetic fields,
    current and magnetic helicity, and shear angles, all of which are confirmed to have strong connections with flares. 

    \item The PIL is important in studying successive flares in the same AR.
    We do see more significant changes in the selected parameters in the PIL area, in particular during big flares.
    A significant store and release process of the free energy before and after flares can be found in both the PIL area and the whole AR, 
    but it is more concentrated in the PIL area. 
    The horizontal field becomes more dominant after flare eruptions in the PIL area, and the magnetic field is also more twisted in the PIL area.
    However, during the evolution of an AR, the coverage and number of PILs may vary significantly.
    This would increase the difficulty of investigating the mechanisms that drive successive solar eruptions.

    \item The establishment of a practical flare forecast system requires full understanding of the physical mechanism of flare bursts.
    A time difference between the parameters and the flare indexes may help us to explore the trigger of flares.
    We have made an example with time difference of two hours, which seems promising, but more work is needed in the future.
\end{enumerate}

A lot of work has been put into identifying the parameters that carry the most information for flare bursts in order to lay foundation for a practical prediction scheme.
\cite{Leka_4} conducted a study on distinguishing the photospheric magnetic field properties that are important for flare production.
Their work shows that the properties related to free energy, total unsigned flux, vertical currents and shear angles are strongly related to flare production.
\cite{bobra2015solar} found that the magnetic energy, vertical current and current helicity 
are the most useful physical quantities in flare prediction, in agreement with \cite{2003Photospheric} and our result here.
\cite{ahmed2013solar} employed a Solar Monitor Active Region Tracker (SMART) to estimate the efficiency of 
such parameters in flare prediction, and selected 6 physical quantities, which are believed to 
carry the most information for flare bursting, out of 21 parameters. 
Unlike the previous studies, we focus on successive flare bursts in the same AR, and identify the most flare-related
SHARP parameters with a time difference of 2 hours.
It still remains a problem to give a practical flare forecasts in the continuously evolving ARs with a high accuracy. 
Interactions between adjacent flares have highly increased the difficulty of establishing a precise prediction method. 
More work about the physical mechanisms of the flares is needed to establish a truly reliable flare forecast system.

\section*{Acknowledgements}
This research was supported by the National Key R$\&$D Program of China No.2021YFA0718600, 
NSFC (41774179, 12073032, 42004145, 42150105),  
the Strategic Priority Research Program of Chinese Academy of Sciences (XDA15018500), 
and the Specialized Research Fund for State Key Laboratories.
Y.G. was supported by NSFC (11773016, 11961131002, and 11533005) and 2020YFC2201201.
H.R. acknowledges helpful discussions with Dr. Huidong Hu.
We acknowledge the work of the Sunpy team \citep{sunpy_community2020}.
We are grateful to the work of \cite{Glogowski2019}, whose Python package \emph{drms} has offered us a convenient tool 
to access the SDO data.
The data used here are courtesy of NASA/SDO and the HMI science team, as well as the Geostationary Satellite System (\emph{GOES}) team.

\appendix
\section*{The NGBoost method}
NGBoost is a gradient boosting algorithm that aims at estimating the parameters of a probability distribution $P_{\theta}(y | \bm{x})$, where $\bm{x}$ is the feature vector of 
observation, and $y$ is the target variable.
In our work, $\bm{x}$ is the vector of SHARP parameters and $y$ is the corresponding flare envelope value. 
NGBoost can also be applied to point estimation by extracting point expectations $\widehat{\mathbb{E}} [y|x]$ from the estimated distribution $ \widehat{P}_{\theta} (y|x)$ \citep{2019NGBoost}.
There are three main modules in NGBoost: (i) the base learner $f$, (ii) the probability distribution $P_{\theta}$, and (iii) a proper scoring rule.
Figure 2 in \cite{2019NGBoost} shows the schematic illustration of NGBoost.
The base learners are crucial in gradient boosting algorithms.
In general, gradient boosting algorithms rely on the sequential training of the base learners to form an additive ensemble. 
The ensemble of the previous learners estimates a current residual, which will then be applied to optimize the new learner. 
The output of that learner is scaled by a learning rate, and it is then appended to the current ensemble.
In our work, we employ the decision tree that is suggested in \cite{2019NGBoost} as the base learner.

Regarding the probability distribution, we choose the normal probability distribution.
It is defined by $\theta$, which is a vector composed of ``location'' and ``scale''.
The next step is to determine a proper scoring rule. 
A proper scoring rule generates a rating between the estimated distribution and the observation, then returns the rating to the base learner such that the distribution of the outcomes gets the best score in expectation.
The most commonly used scoring rule, which is also applied in this context, is the logarithmic score:
\begin{equation}
    \mathcal{L} (\theta, y) = -log(P_{\theta} (y)).
\end{equation}
Another important part of NGBoost is the natural gradient $\tilde{\bigtriangledown}_{\theta}$.
The natural gradient is motivated by information geometry. 
It makes the optimization problem invariant to the parameterization, 
and it also makes the learning more stable and effective \citep{amari1998natural}.

\begin{figure}
    \centering
    \includegraphics[width=1.0\textwidth, height=0.40\textwidth]{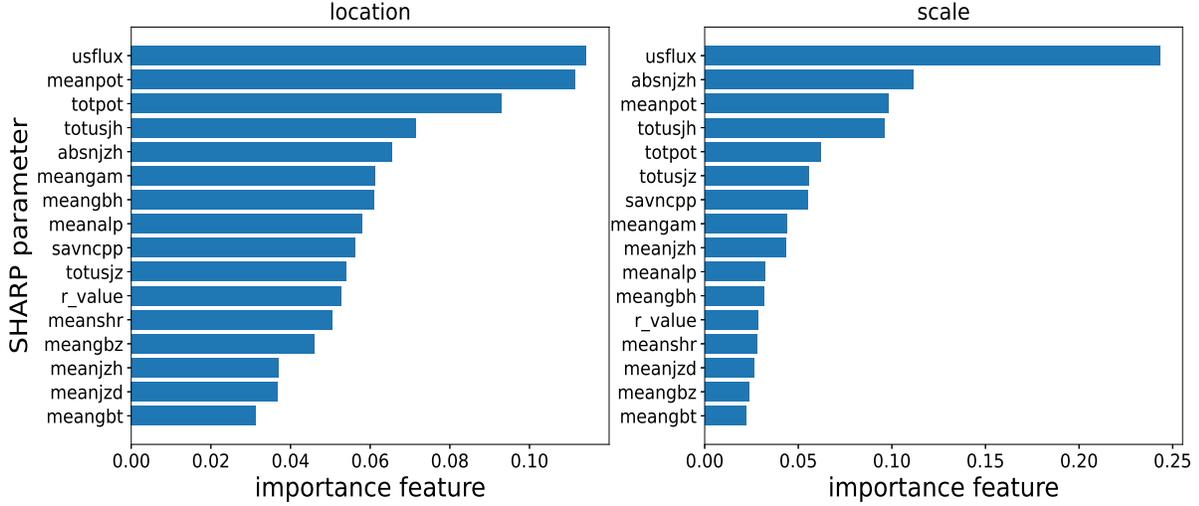}
    \caption{The importance feature of the SHARP parameters in AR 12673. 
    The left panel refers to the importance feature distribution of ``location'', and the right panel refers to the importance feature distribution of ``scale''.}
    \label{fig:importance_bar}
\end{figure}

To be specific, assuming that we have a dataset $D = \{x_i, y_i\}_{i=1}^M$, where $M$ is the number of the training examples.
The algorithm first establishes an initial predicted parameter $\theta^{(0)}$, which is defined as:
\begin{equation}
    \theta^{(0)} = argmin_{\theta}\sum_{i=1}^M\mathcal{L} (\theta, y_i).
\end{equation} 
The initial predicted parameter is a standard set of parameters that aims at minimizing the sum of the scoring rule $\mathcal{L}$.
Next, for each iteration n, we estimate the individual natural gradient for each example i as:
\begin{equation}
    g_i^{(n)} = \mathcal{I}_{\mathcal{L}}(\theta_i^{(n-1)})^{-1} \cdot \bigtriangledown_{\theta} \mathcal{L} (\theta_i^{(n-1)}, y_i),
\end{equation}
where the $\mathcal{I}_{\mathcal{L}} (\theta)$ is the Fisher Information carried by an observation about $P_{\theta}$. For logarithmic score, the Fisher Information is defined as:
\begin{equation}
    \mathcal{I}_{\mathcal{L}} (\theta) = \mathbb{E}_{y \thicksim P_{\theta}} [\bigtriangledown_{\theta} \mathcal{L} (\theta, y) \bigtriangledown_{\theta} \mathcal{L} (\theta, y)^T].
\end{equation}
After being established, the natural gradient and the input vectors $\pmb{x}_i$ are used to train the base learners of iteration $f^{(n)}$. The training process returns a fitted base learner that is the projection of the natural gradient.
A scale factor $ \rho^{(n)} $ is needed to scale the projected gradient. It is computed as:
\begin{equation}
    \rho^{(n)} = argmin_{\rho} \sum_{i=1}^{M} \mathcal{L} (\theta_i^{(n-1)} - \rho \cdot f^{(n)} (\bm{x}_i), y_i),
\end{equation}
where $\rho$ starts from $\rho=1$.
Then the parameters $\theta_i^{(n)}$ are updated by:
\begin{equation}
    \theta_i^{(n)} = \theta_i^{(n-1)} - \eta (\rho^{(n)} f^{(n)} (\bm{x}_i)),
\end{equation}
where $\eta$ denotes a usual learning rate (typically 0.1 or 0.01).

The use of NGBoost in this context is explained in Section 2.4. 
We focus on the importance feature of each SHARP parameter. 
The SHARP parameter that has higher importance feature value in the NGBoost analysis is believed to be more responsible for successive flare occurrence. 
An example of the importance feature distribution is shown in Figure \ref{fig:importance_bar}.
The NGBoost method can be conveniently used and interpreted \footnote{https://stanfordmlgroup.github.io/ngboost/intro.html} 
with the python package \emph{ngboost}\footnote{https://github.com/stanfordmlgroup/ngboost}.

\newpage
\bibliography{sample631}{}
\bibliographystyle{aasjournal}

\end{document}